\documentclass[twocolumn,prb,showpacs,preprintnumbers,amsmath,amssymb,superscriptaddress]{revtex4}

\usepackage{subfigure}
\usepackage{graphicx}
\usepackage{dcolumn}
\usepackage{bm}
\usepackage{epsfig}

\begin{document}


\title{The malleability of uranium: manipulating the charge-density wave in epitaxial films}
\author{R. Springell}
\affiliation{Royal Commission for the Exhibition of 1851 Research Fellow, Interface Analysis Centre, School of Physics, University of Bristol, UK, BS2 8BS, UK}
\author{R. C. C. Ward}
\affiliation{Clarendon Laboratory, University of Oxford, Oxford, Oxon OX1 3PU, UK }
\author{J. Bouchet}
\affiliation{CEA, DAM, DIF, F-91297 Arpajon, France}
\author{J. Chivall}
\affiliation{Department of Physics and Astronomy, University College London, London WC1E 6BT, UK}
\author{D. Wermeille}
\altaffiliation{Department of Physics, University of Liverpool, Liverpool L69 7ZE, UK}
\affiliation{XMaS, European Synchrotron Radiation Facility, BP220, F-38043 Grenoble Cedex 09, France}
\author{P. S. Normile}
\altaffiliation[Present address: ]{Universidad de Castilla-La Mancha, Facultad de Ciencias Químicas, Avda. Camilo José Cela s/n, 13071 Ciudad Real, Spain}
\affiliation{XMaS, European Synchrotron Radiation Facility, BP220, F-38043 Grenoble Cedex 09, France}
\author{S. Langridge}
\affiliation{ISIS, Rutherford Appleton Laboratory, Chilton, Oxfordshire OX11 0QX, UK}
\author{S. W. Zochowski}
\affiliation{Department of Physics and Astronomy, University College London, London WC1E 6BT, UK}
\author{G. H. Lander}
\affiliation{European Commission, Joint Research Centre, Institute for Transuranium Elements, Postfach 2340, D-76125 Karlsruhe, Germany}


\begin{abstract}
We report x-ray synchrotron experiments on epitaxial films of uranium, deposited on niobium and tungsten seed layers. Despite similar lattice parameters for these refractory metals, the uranium epitaxial arrangements are different and the strains propagated along the \textit{a}-axis of the uranium layers are of opposite sign. At low temperatures these changes in epitaxy result in dramatic modifications to the behavior of the charge-density wave in uranium. The differences are explained with the current theory for the electron-phonon coupling in the uranium lattice. Our results emphasize the intriguing possibilities of producing epitaxial films of elements that have complex structures like the light actinides uranium to plutonium.
\end{abstract}

\pacs{75.80.+q, 77.65.-j}

\maketitle

\section{Introduction}

Almost all metallic elements have simple crystal structures (\textit{fcc}, \textit{bcc}, \textit{hcp}, \textit{dhcp}) at ambient pressure and temperature \cite{Donohue}. There are exceptions, of course, such as Mn and Hg, but the most exotic structures are found with the early actinides Pa, U, Np, and Pu. These latter four elements all adopt complicated structures, most have many allotropes before melting, and the ambient structures are all different \cite{Moore}. This structural diversity arises because of the interplay between the partially occupied 5\textit{f} and 6\textit{d} states \cite{Soderlind} and at the nanoscale, these elements may well prove more malleable in forming unexpected epitaxial structures than the more conventional elements.

Epitaxial engineering \cite{Bland} (i.e. the production of thin films in single-crystal form, on atomically ordered substrates) has been practiced now for almost 50 years via various processes, and has illustrated in many ways how the elements can be manipulated, principally through "lattice matching", thus reducing the interfacial strain between substrate and film.  However, the simplicity of the atomic structures of most elements constrains the available options. Chromium, for example, has fascinating properties, such as the spin-density wave (SDW). Changes in the SDW, induced by epitaxial engineering, are significant \cite{Zabel}, but are restricted by the isotropy and robustness of the underlying \textit{bcc} lattice of Cr. With the light actinides, however, the structural diversity implies the possibility of many new effects and structures that are not observed in the bulk. We have shown earlier \cite{Springell} that \textit{hcp}-U films can be stabilized, a structure that cannot be found in the bulk phase diagram. At elevated temperature (above 1050\,K) uranium exists in a \textit{bcc} structure - if this could be stabilized at low temperature it might order magnetically, as a consequence of the large inter U spacing. Similarly, plutonium exhibits both the \textit{fcc} ($\delta$-phase) and \textit{bcc} ($\varepsilon$-phase) above $\sim$580\,K and $\sim$870\,K, respectively. To our knowledge, epitaxial films of such transuranic elements are yet to be synthesized, and this represents a challenge for the future. Our results form the first step in such a task.

Alpha-uranium (the stable crystal structure at ambient pressure and temperature) is famous for being the only element that spontaneously exhibits a charge-density wave (CDW), which occurs at $T_{0}$\,=\,43\,K \cite{Lander}.  Recently, the CDW has been investigated in more detail, both theoretically \cite{Bouchet} and experimentally \cite{Raymond}, emphasizing the importance of the strong electron-phonon coupling along the [100] axis. The results show that the length of this [100] $a_{U}$-axis is the key parameter in determining the behavior of the CDW. Furthermore, as the CDW is suppressed by pressure, the temperature at which uranium becomes superconducting increases \cite{Lander}, demonstrating the link between the two phemonena, as shown recently in high-$T_{C}$ materials \cite{Chang}, and placing uranium in the context of such materials of interest from a fundamental perspective.

Earlier, we reported geometric relationships at room temperature for the orthorhombic (space group Cmcm) $\alpha$-U structure with the [110] growth axis on Nb, and the [001] axis on W \cite{Ward}. In the present work we demonstrate how the malleability of uranium allows it to form different epitaxial structures with these two commonly used buffer materials, Nb and W, and that the strains produced for the two orientations on the important [100] uranium axis gives rise to very different behaviors of the resulting CDW's.

\section{Epitaxy conditions}

\begin{figure}
  {\includegraphics[width=0.4\textwidth,bb=100 40 570 440,clip]{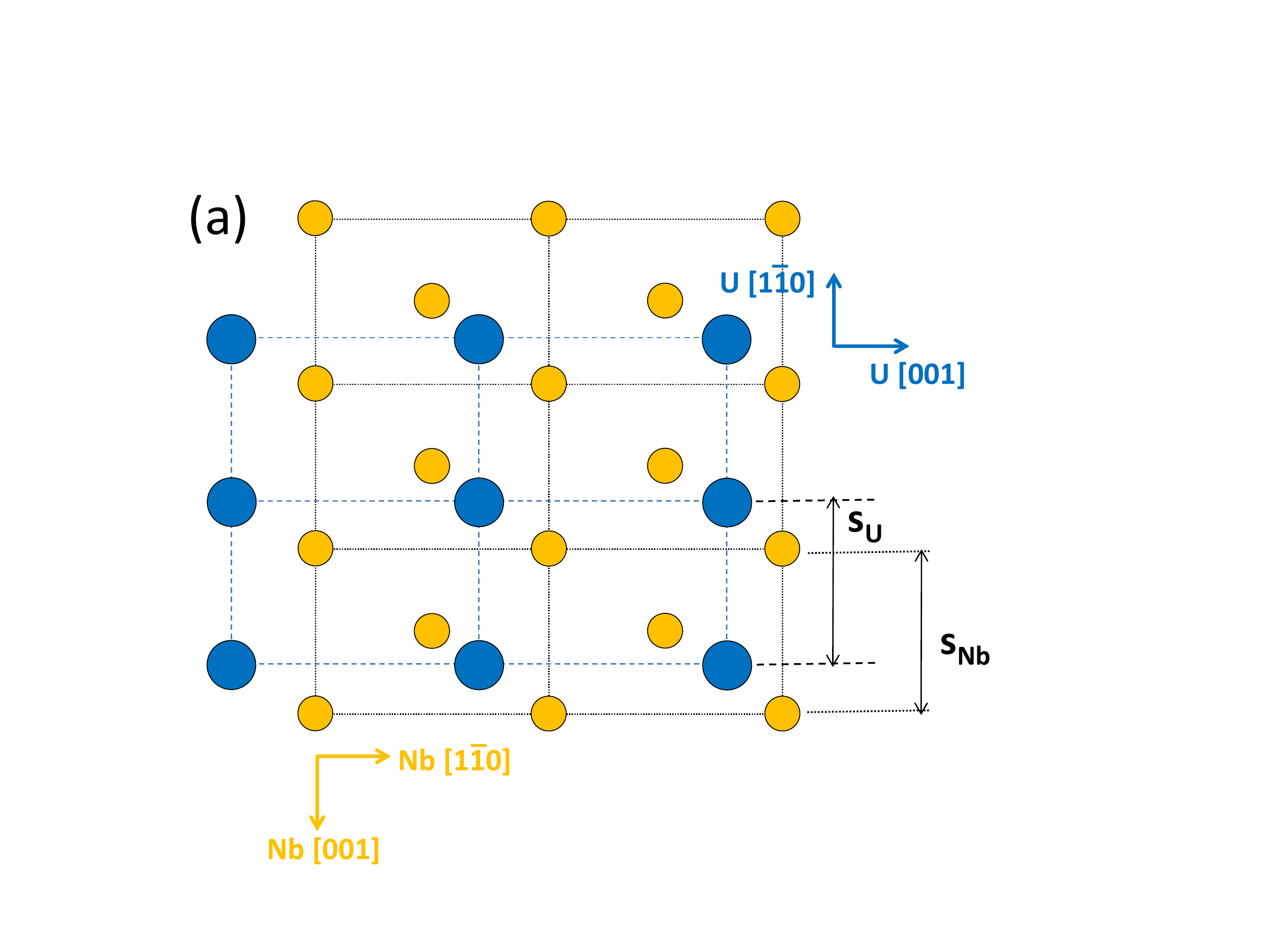}} \\
  {\includegraphics[width=0.4\textwidth,bb=90 80 590 540,clip]{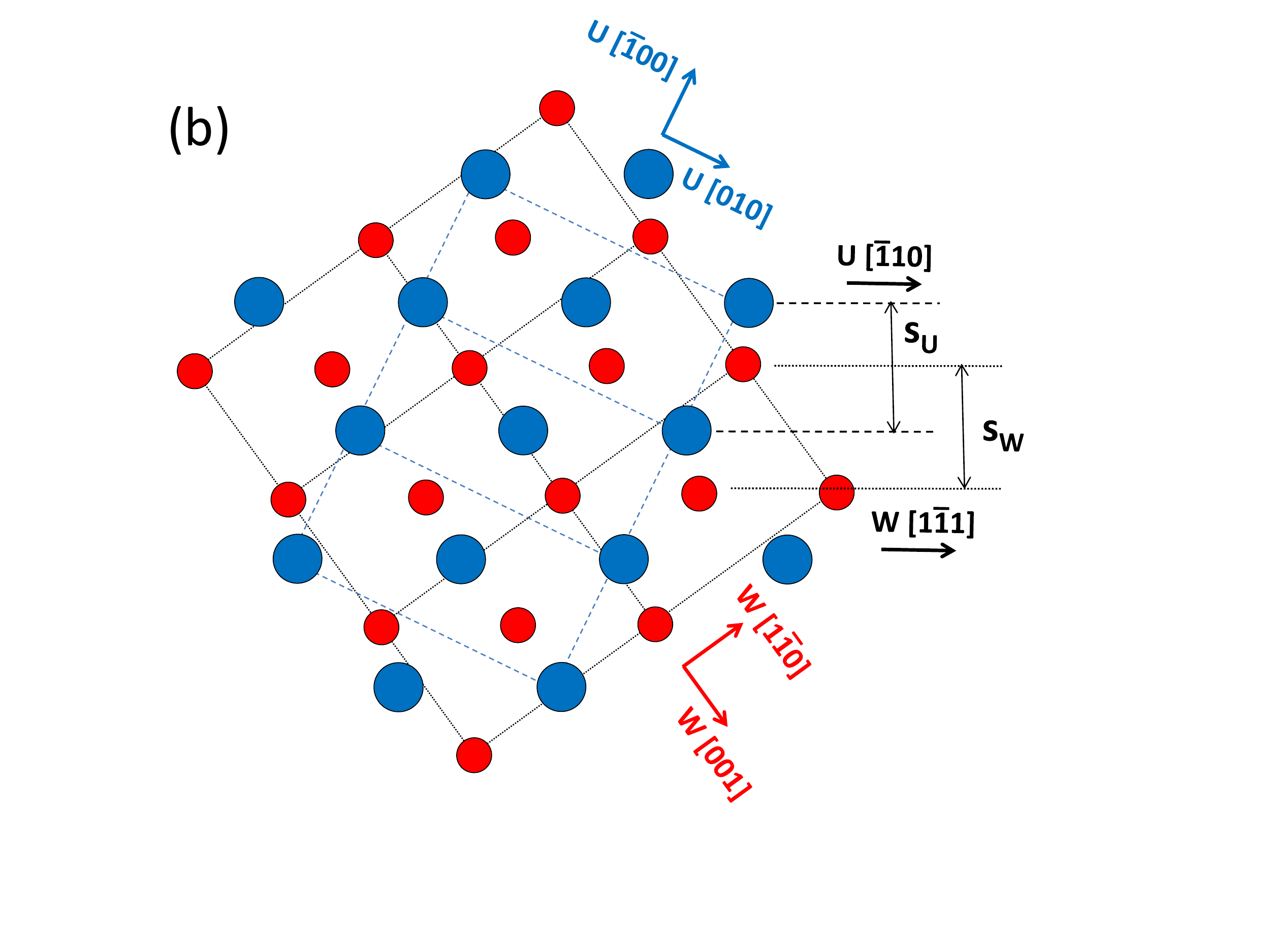}}
  \caption{Epitaxial relationship for $\alpha$-U (blue circles) on (a) Nb ($a_{0}=3.300\,\AA$) and (b) on W ($a_{0}=3.165\,\AA$). For the (a) orientation the governing factor is the distance in the horizontal plane between rows of U atoms, $s_{U}=\frac{1}{2}(a_{U}^{2}+b_{U}^{2})^{\frac{1}{2}}$=3.264\,\AA. For the orientation, (b), the rows of U and W atoms must be within register so it is necessary that the uranium $s_{U}=d(110)_{U}$ must be close to the tungsten $s_{W}=2\times d(112)_{W}$.}
  \label{fig:1}
\end{figure}

Figure \ref{fig:1} shows the epitaxial relationships reported by Ward \textit{et al.} \cite{Ward} for U grown on the refractory metal buffers Nb(110) and W(110), deposited on (11.0) plane sapphire substrates. Although the difference between the lattice parameters of Nb and W is only 4.3\%, the orientations that $\alpha$-U adopts for epitaxy on these two elements are different.

On Nb(110), $\alpha$-U grows in a (110) orientation and the epitaxy is governed by the fit between the U[1-10] and Nb[001] rows of atoms in the interfacial plane, i.e. the horizontal atomic rows in Fig. \ref{fig:1}(a). The calculated misfit ($\mathrm{\Delta=½(s_{U}-s_{Nb})/(s_{U}+s_{Nb})}$) is -1.1\% at room temperature, and increases slightly (-1.4\%) at the growth temperature (T$_{d}$) of 450$^{\circ}$C . Note that the misfit in the perpendicular in-plane direction ([001] of $\alpha$-U) is much larger (+6.2\%), but this is a common feature of metal epitaxy, where a match in one direction between parallel, close-packed rows of atoms at the interface is often the governing factor.

In the case of W(110) the corresponding misfit for the epitaxy of Fig. \ref{fig:1}(a) is +2.8\%. This is too large to be acceptable, and instead the $\alpha$-U prefers to grow in the (001) orientation, as shown in Fig. \ref{fig:1}(b). In this case the in-plane parallel rows of atoms are U[-110] and W[1-11], which have a misfit in spacing of only +0.2\% at the growth temperature. By comparison, the corresponding misfit for U/Nb would be -4.1\% which is unacceptably large. A feature of low-symmetry structures such as orthorhombic $\alpha$-U is that there exist many more optional orientations available for epitaxy, and the lowest-energy relationships are often difficult to predict.

The final strains found in the U layers depend not only on lattice mismatch but also the substrate clamping effect due to the different thermal expansion coefficients of substrate and layers. This latter effect is particularly significant in our case because of the large and anisotropic linear thermal-expansion coefficients ($\alpha$) of uranium \cite{Lloyd}. At T$_{d}$ the values are $\alpha_{U}$[100]=+33, $\alpha_{U}$[010]=-6.1, $\alpha_{U}$[001]=+30.6 (all in units $\times10^{-6}$\,K$^{-1}$). In contrast, for the refractory \textit{bcc} metals and the sapphire substrate, the $\alpha$ coefficients show little temperature dependence and are all between 5 and 9 $\times10^{-6}$\,K$^{-1}$. The substrate clamping effect introduces a crucial difference in the state of strain of the U(001) and U(110) layers. In the case of U(001), both $a_{U}$ and $b_{U}$ are in-plane and are restrained by the substrate from contracting ($a_{U}$) or expanding ($b_{U}$), as they would like on cooling to room temperature. We therefore expect $a_{U}$ to be in tension and $b_{U}$ to be in compression for U/W. Along the growth direction, $c_{U}$ is free to respond to the strained in-plane cell parameters, and is expected to change to preserve the unit-cell volume. On the other hand, for U(110) as in U/Nb only $c_{U}$ is in-plane and will be in tension after cooling; $a_{U}$ is the axis closest to the surface normal and would therefore be expected to be in compression to maintain the atomic volume.

Thus the different U orientations found on the two refractory metal buffers, a feature of the low symmetry of $\alpha$-U, together with its anomalous thermal expansion coefficients, result in the $a_{U}$ axis being in compression on Nb and in tension on W. Because of the importance of $a_{U}$ to the CDW transition, these strains are anticipated to lead to a different behavior of the CDW at low temperature between the U/Nb and U/W samples.

\section{Experimental results}

All experiments have been performed, using a monochromated beam of 10\,keV x-rays at the XMaS beamline (BM28) \cite{XMaS} at the European Synchrotron Radiation Facility, Grenoble. All samples were grown, using a dedicated uranium deposition system, developed at Oxford University and now housed at the University of Bristol \cite{Springell,Ward}.

\subsection{Case of U/Nb}

In the case of U/Nb, as already discussed \cite{Springell} for a 5000\,\AA\, sample, the CDW appears at approximately the same $T_{0}$ as in the bulk (43\,K) and with the same wave-vector components \cite{Lander}. We have examined a large number of epitaxial samples, ranging from 70 to 2000\,\AA \cite{Chivall}, and in all cases the CDW appears at a similar $T_{0}$, with similar components, to those reported in Ref. [6]. A comparison to measurements on bulk samples \cite{Lander}, shows (by normalizing to a lattice peak) that the CDW in U/Nb epitaxial films is reduced in intensity compared to the bulk, and the domain population is heavily biased, unlike in the bulk.

\begin{figure}
  {\includegraphics[width=0.45\textwidth,bb=100 70 580 480,clip]{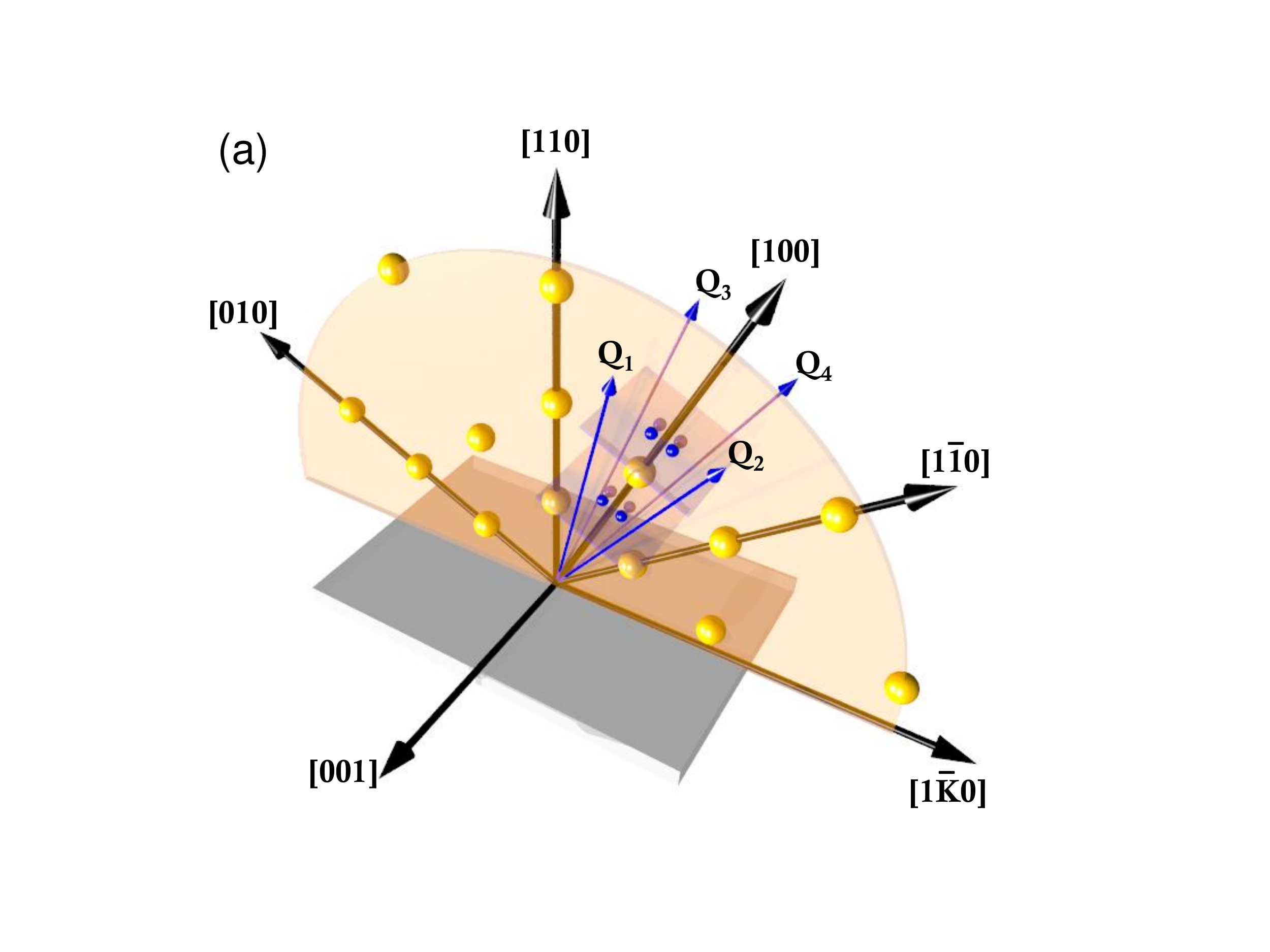}} \\
  {\includegraphics[width=0.45\textwidth,bb=10 0 740 520,clip]{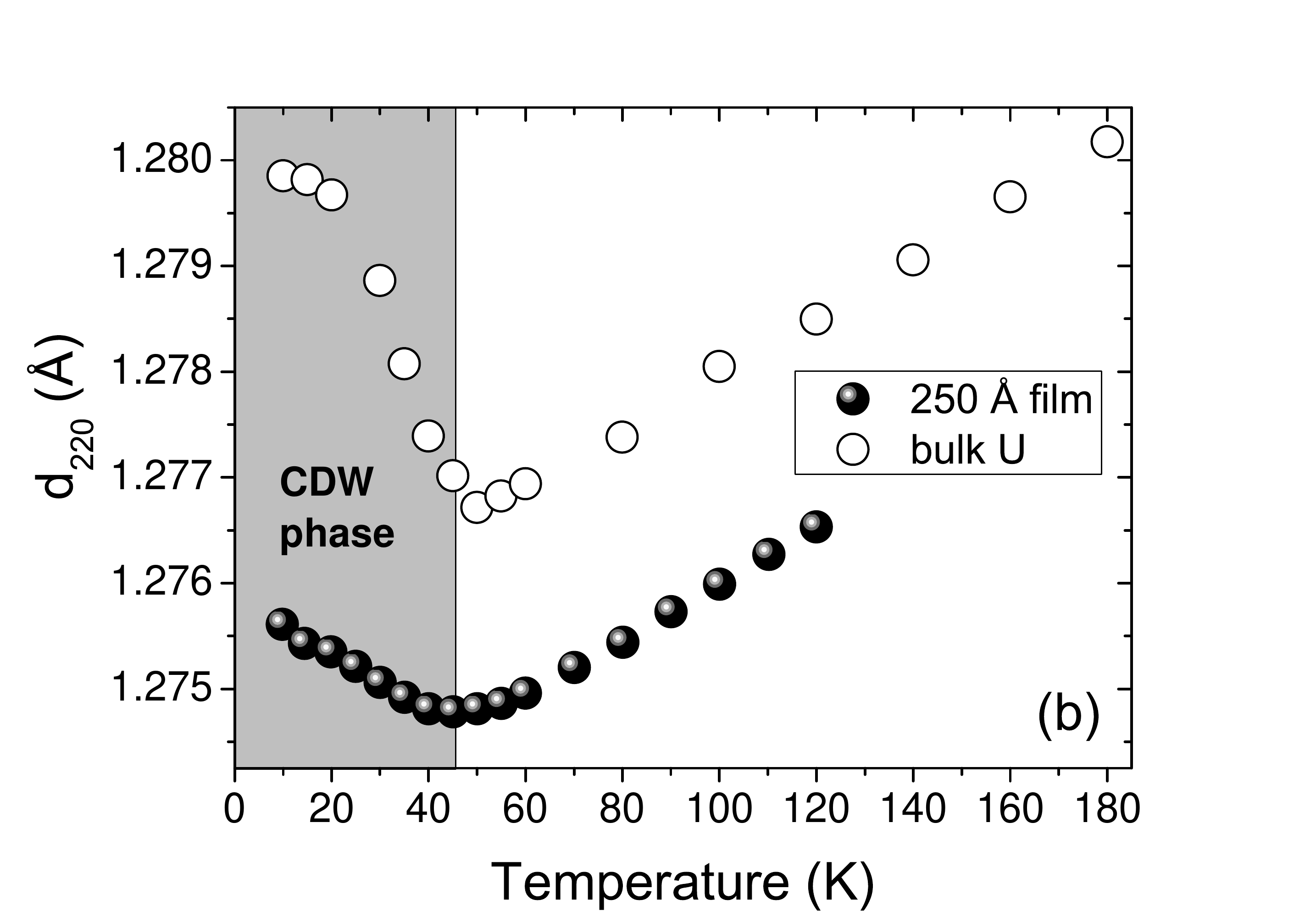}}
  \caption{(a) Plane of the film for U/Nb. The growth direction is [110], with the $a_{U}$ axis at 25.8$^{\circ}$ to this [110] direction. The four CDW domains that can be readily accessed are $\textbf{Q}_{\textbf{1}}\mathrm{(+q_{x} +q_{y} +q_{z})}$, $\textbf{Q}_{\textbf{2}}\mathrm{(+q_{x} -q_{y} +q_{z})}$, $\textbf{Q}_{\textbf{3}}\mathrm{(+q_{x} +q_{y} -q_{z})}$, and $\textbf{Q}_{\textbf{4}}\mathrm{(+q_{x} -q_{y} -q_{z})}$ where $\mathrm{q_{x}=0.5\,a^{\ast}}$, $\mathrm{q_{y}=0.22\,b^{\ast}}$ and $\mathrm{q_{z}=0.167\,c^{\ast}}$ are symmetrically spaced at 17$^{\circ}$ away from the $a_{U}$ axis [100]. (b) Lattice spacing $d(220)_{\mathrm{U}}$ as a function of temperature for both a 250\,\AA-thick U/Nb film (solid circles) and a bulk sample (open circles). The temperature at which the CDW develops ($T_{0}$) is 45\,K, as in bulk $\alpha$-U.}
  \label{fig:2}
\end{figure}

In these films, as shown in Fig. \ref{fig:2}(a), the $a_{U}$ axis, [100], is marked, as are the four CDW wave-vectors, $\textbf{Q}_{\textbf{1}}$, $\textbf{Q}_{\textbf{2}}$, $\textbf{Q}_{\textbf{3}}$, and $\textbf{Q}_{\textbf{4}}$. Since $\pm$\,$q_{z}$ are equal displacements with respect to the film, we expect the domain intensities of $\textbf{Q}_{\textbf{1}}$, and $\textbf{Q}_{\textbf{3}}$, on one hand, and $\textbf{Q}_{\textbf{2}}$, and $\textbf{Q}_{\textbf{4}}$, on the other, to be equivalent. This is experimentally found - see Fig. 2 of Ref. [6] - in all films. However, domain $\textbf{Q}_{\textbf{1}}$ is found to have at least 100 times the intensity of domain $\textbf{Q}_{\textbf{2}}$. This imbalance appears because $\textbf{Q}_{\textbf{2}}$ has a larger component in the plane of the film than for $\textbf{Q}_{\textbf{1}}$. The CDW thus favors domain $\textbf{Q}_{\textbf{1}}$ as the in-plane ($a_{U}$ and $b_{U}$) axes are subject to less clamping from the buffer and substrate than in domain $\textbf{Q}_{\textbf{2}}$.

Although the CDW satellite peaks give directly the ordering temperature $T_{0}$ (and periodicity) of the CDW, it is also instructive to examine the lattice peaks as a function of temperature. In Fig. \ref{fig:2}(b) we show the d(220)$_{U}$ plane spacing from a 250\,\AA-thick epitaxial film of U/Nb. Similar figures exist for all samples. These measure the spacing of the atomic planes perpendicular to the [110] growth direction. The bulk values are taken from Barrett \textit{et al.} \cite{Barrett}. The film value is slightly smaller than the bulk one, consistent with the compression, as discussed above in the U/Nb configuration, and the relative change of the d(220)$_{U}$ plane spacing below $T_{0}$ is far less than that found in the bulk. All these features, as well as the domain imbalance, are consistent with the compression of the $a_{U}$ axis in the U/Nb films.

\subsection{Case of U/W}

The epitaxy of U/W is as shown in Fig. \ref{fig:1}(b) with the $a_{U}$ and $b_{U}$ axes in the plane of the film, and the growth direction [001]. A 1500\,\AA\, film exhibits a rocking curve (full width at half maximum) of 0.35$^{\circ}$. The only difference in the epitaxial relationship of this film with those discussed in Ref. [11] is that we have deposited a thin (100\,\AA) seed layer of Nb on top of the sapphire substrate before depositing the 250\,\AA\, buffer of W. This reduces the number of domains of the W buffer, from two to one. When the uranium is deposited, the number of domains is then reduced from four to two (B1 and B4 in Fig. 7 of Ref. [11]), whereas in these earlier studies \cite{Ward} up to six domains were reported.

\begin{figure}
\includegraphics[width=0.4\textwidth,bb=120 110 360 410,clip]{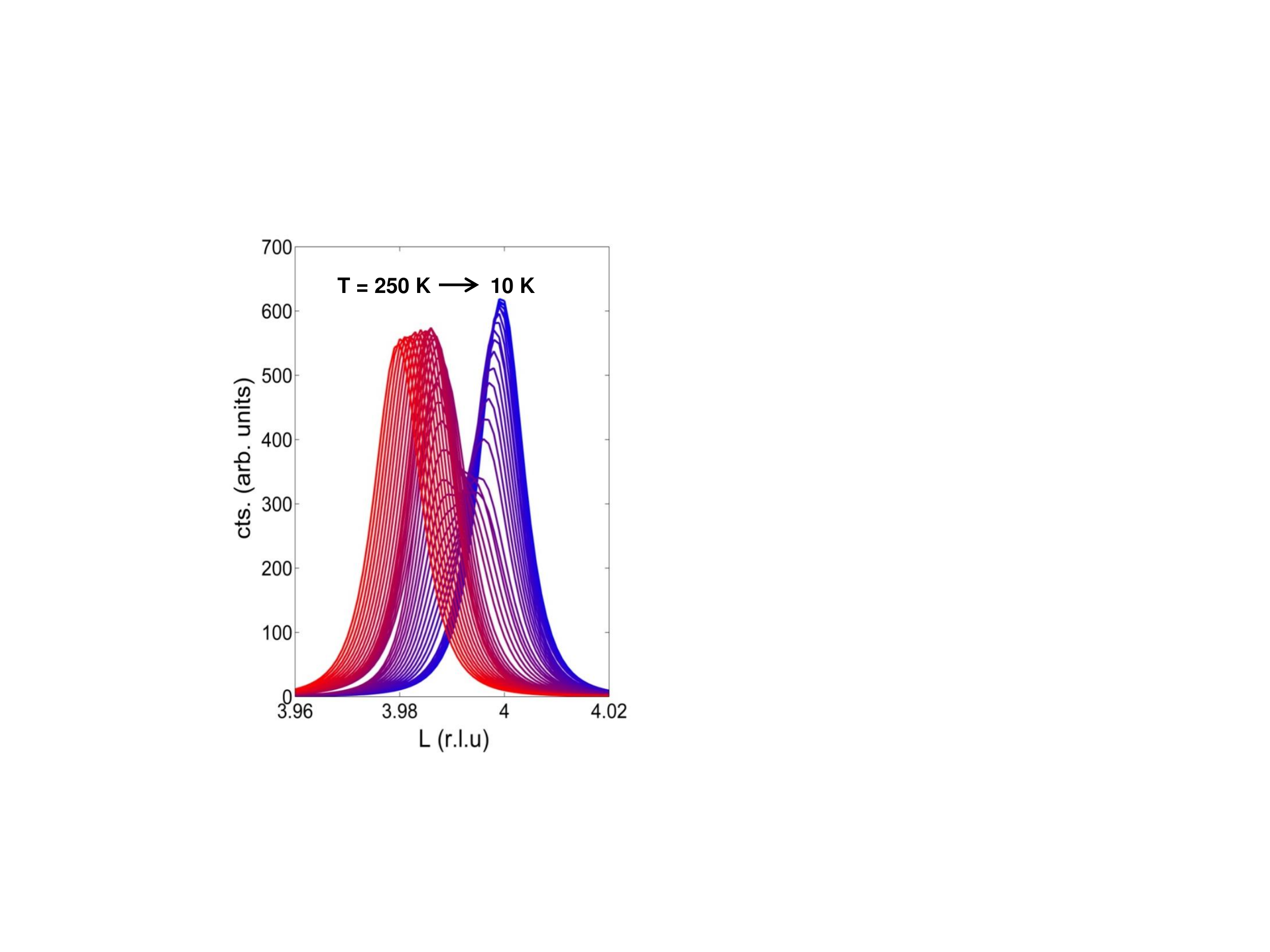}\caption{The curves, which represent interpolation between data points, show the high-temperature lattice parameter (red) varying only slowly with temperature for 300\,$>$\,T\,$>$\,150\,K, and then the emergence of a new (smaller) $c_{U}$ lattice parameter for T\,$<$\,150\,K (blue). The indexing is normalized to L\,=\,4 at base temperature. The curves can be fitted with identical widths ($\Delta$L/L) for the two different lattice parameters at all temperatures.\label{figure3}}
\end{figure}

Since the change in $c_{U}$ can be gauged directly from the position of the (004) reflection, we show this in Fig. \ref{figure3} as a function of temperature. Curiously, the widths of the (004) reflection, which are a measure of the correlations across the thickness of the film in the [001] direction, remain independent of temperature. Thus, the domains in the [001] direction are transforming from one lattice parameter to the other; they do not co-exist in the same domain, as this would give rise to a broadening of the peaks. The widths are the same for the high- and low-temperature phases, both reflecting the finite thickness of the film. The width of off-specular Bragg reflections with a greater in-plane component, shows a consistent broadening, indicating that the lateral dimension of the structural domains is even smaller than the film thickness of 1500\,\AA.

\begin{figure}
\includegraphics[width=0.45\textwidth,bb=0 0 430 860,clip]{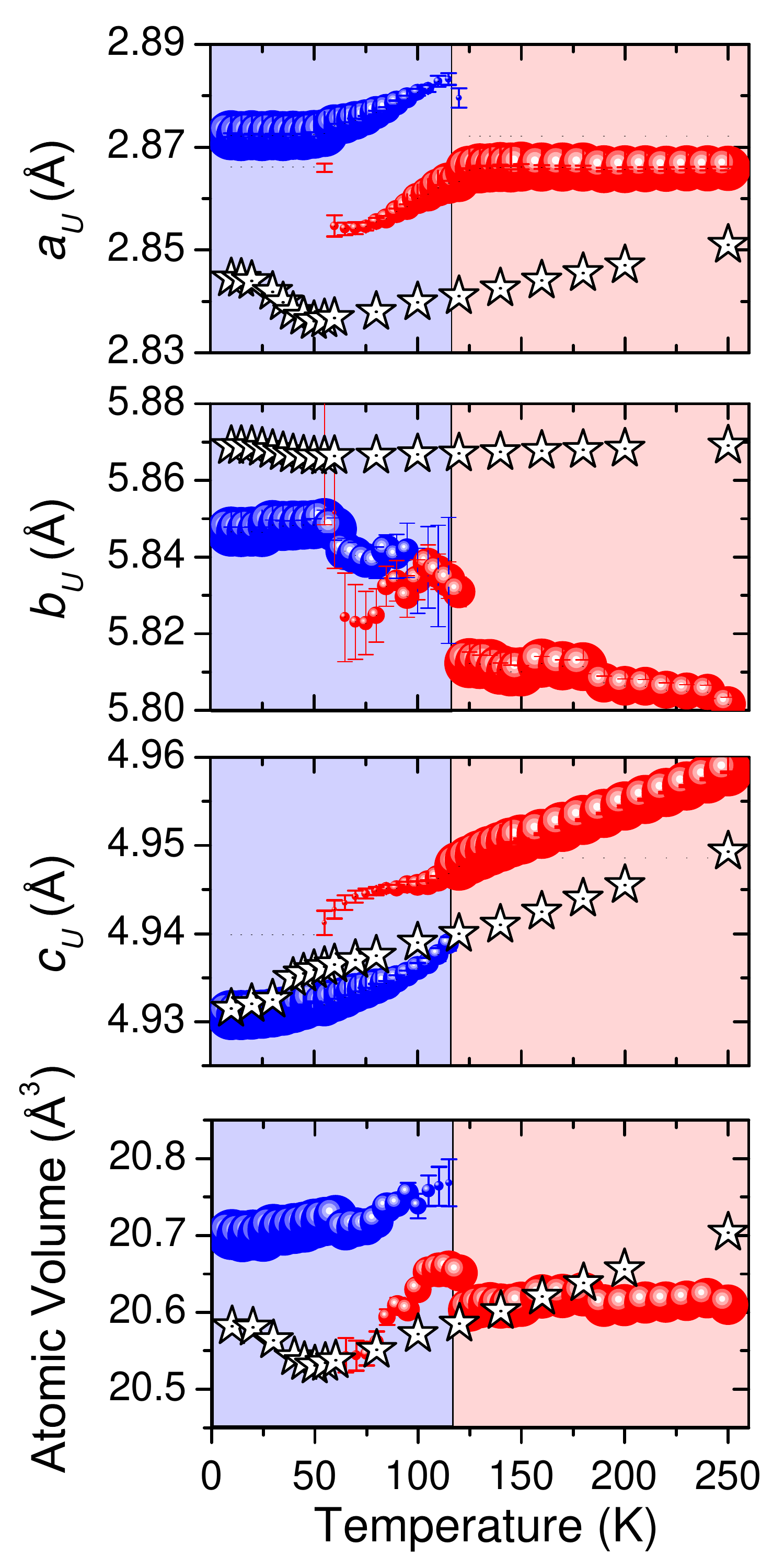}\caption{$a_{U}$, $b_{U}$, and $c_{U}$ lattice parameters, and the atomic volume as a function of temperature for the U/W film of 1500\,\AA. The open stars give the bulk values \cite{Barrett}. The red points represent the high-temperature phase and the blue, the low temperature phase, in which the CDW appears. The size of the points is an indication of the volume of the sample that has such a lattice parameter. \label{figure4}}
\end{figure}

The temperature dependence of the $a_{U}$, $b_{U}$, and $c_{U}$ lattice parameters are shown in Fig. \ref{figure4}. Recall that the growth direction is [001] $c_{U}$, so that the changes in this parameter act to preserve the atomic volume. The largest change (of almost 1\%) is in $a_{U}$. Initially, from ambient to $\sim$150\,K, contraction over this temperature range of both buffer and substrate are unobservable on this scale. However, as $a_{U}$ starts to contract at lower temperatures, a new, larger, $a_{U}$ emerges, and by the lowest temperature the complete volume of the film exhibits this new $a_{U}$. The atomic volume (see bottom panel of Fig. \ref{figure4}) of the U/W film matches that of the bulk at ambient temperature, but at low-temperature is 0.5\% larger, reflecting mainly the expansion of the $a_{U}$ axis.

\begin{figure}
\includegraphics[width=0.5\textwidth,bb=0 230 580 560,clip]{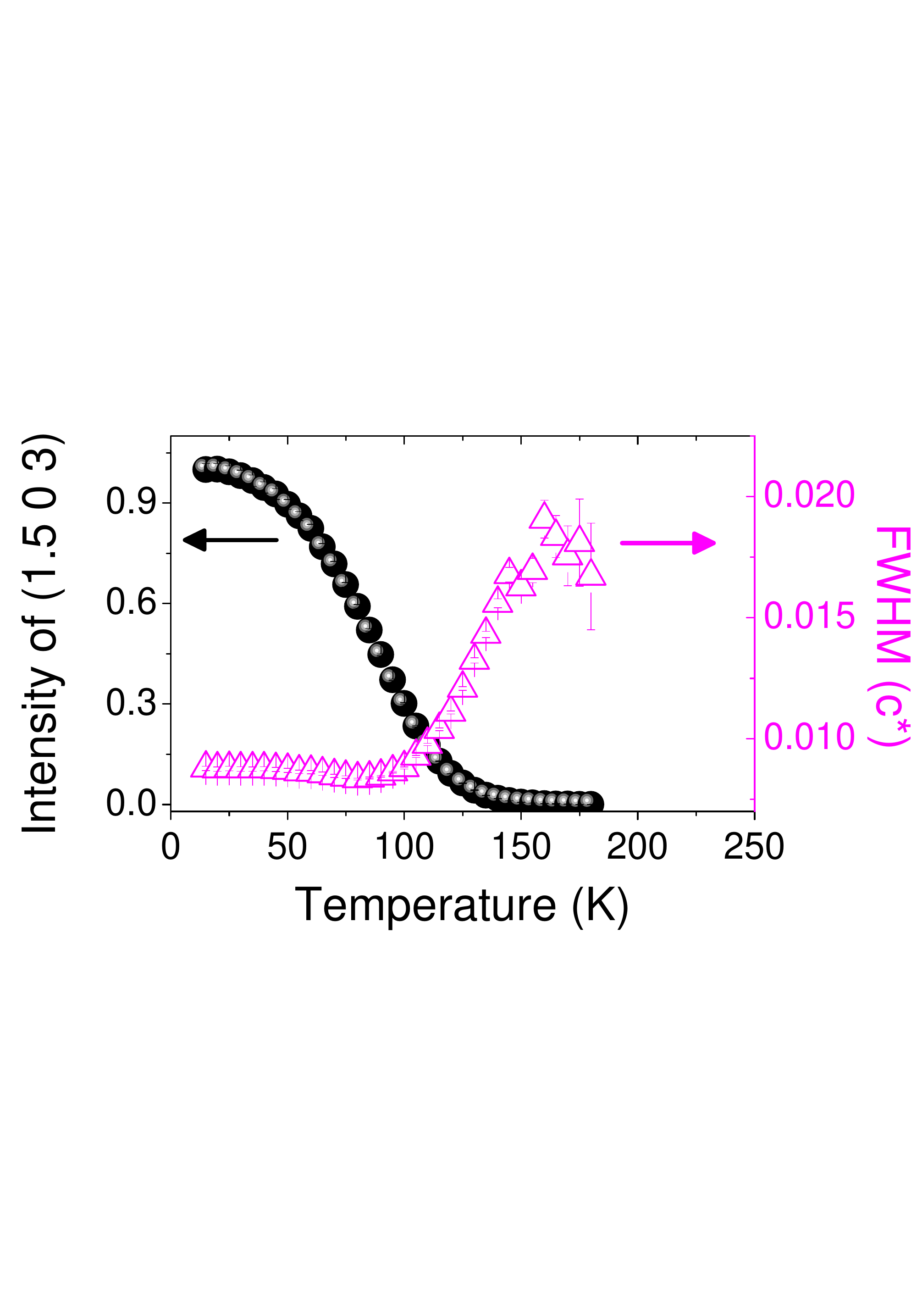}\caption{The integrated intensity of the (1.5 0 3) CDW reflection as a function of temperature for the 1500\,\AA film is represented by the solid black points and the left-hand y-axis scale. The width of the (1.5 0 3) reflection is represented by the open magenta triangles and the right-hand y-axis scale.
\label{figure5}}
\end{figure}

Accompanying the large change in $a_{U}$ at low temperature is the appearance of a new set of satellites ($q_{x}=0.5, q_{y}=q_{z}=0$), and the temperature dependence of the (1.5 0 3) reflection is shown in Fig. \ref{figure5}, together with its full-width at half maximum. Since this satellite has no $q_{y}$ or $q_{z}$ components, the CDW is different from that found in bulk $\alpha$-U, but closely related. It is the so-called $\alpha_{1}$-phase, as discussed in Ref. [8], and incorporates the principal physics of the CDW in terms of the strong electron-phonon interaction, which is known \cite{Raymond} to have its maximum amplitude at the position (0.5 0 0) in the Brillouin zone. $T_{0}$ is now 120\,K, rather than the bulk value of 43\,K, an almost three-fold increase. We observe diffuse scattering corresponding to the soft phonon at this position (see below) up to $\sim$180\,K, and the width of this scattering increases as a function of $\Delta$T from $T_{0}$, as expected for a phonon-mediated phase transition \cite{Raymond,Marmeggi}. An estimate of the $\beta$-value for the growth of the intensity of the CDW peak gives 0.53$\pm$0.03, consistent with a simple Landau-type order parameter, as suggested by earlier work on the soft-phonon that drives this transition \cite{Marmeggi}. The width of the CDW peak, measured in the [001] direction, is approximately 0.009 r.l.u, which corresponds closely to that found for the (004) charge peak, as in Fig. \ref{figure3}. Thus, the CDW extends across the whole film thickness, however, above $T_{0}$ much shorter-range correlations exist.

At a lower temperature ($\sim$45\,K) we observe the small incommensurate satellites along the $b_{U}^{\ast}$ and $c_{U}^{\ast}$ reciprocal axes corresponding to the $\textbf{Q}_{\textbf{CDW}}$ described for the U/Nb samples (see ref. [6] and above) and found in the bulk \cite{Lander}, but they are very weak ($<$1\% of the main satellites) and almost certainly arise from strain effects \cite{Lander}. What is unique about this U/W epitaxy is that the CDW is formed at much higher temperatures than in the bulk, and it appears with just the $q_{x}=0.5$ component. The peak at (1.5 0 3) is intense ($\sim$5\% of the strong (202) charge reflection) and corresponds to a displacement of the U atoms by $\sim$0.07\,\AA\, from their equilibrium positions, which is more than twice as large as that found in the bulk \cite{Lander}.

\section{Theory}

We now turn to an understanding of the development of the CDW as a function of the electronic structure of uranium. Our \textit{ab-initio} calculations have been performed following Ref. [8] and Ref. [9]. As shown in these previous works, there is an intrinsic soft-mode in the $\alpha$-U structure that is a result of the electron-phonon interaction along the [100] direction, peaked at $h$=0.5, and this drives the formation of the CDW. The $\alpha$-U structure is not stable at T=0\,K, as demonstrated by the results for the bulk (solid blue line) in Figure \ref{figure5}. Instead, the stable structure is the $\alpha_{1}$ structure with a doubling of the $a_{U}$ axis. Similarly, calculations using cell parameters corresponding to the film (solid red line) show, as expected, that the $\alpha$-U structure is even more unstable in the film, and it is not surprising that the $\alpha_{1}$ structure is formed at a higher $T_{0}$ than found in the bulk. The inset shows the changes in the soft-phonon energy as a function of changes in the $a_{U}$ parameter.

\begin{figure}
\includegraphics[width=0.45\textwidth,bb=30 20 460 550,clip]{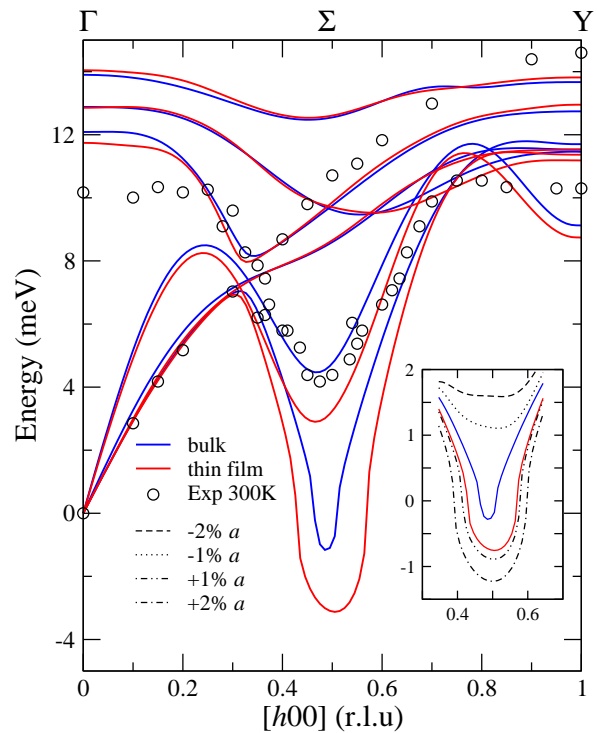}\caption{Theoretical results using the same calculations as reported in Ref. [8]. The experimental points at room temperature for key modes are shown - see Ref. [7]. The $\alpha$-U structure is unstable at T\,=\,0\,K as shown by the soft phonon at $h$\,=\,0.5 having an imaginary frequency. The inset shows how this frequency changes as $a_{U}$ is either compressed or expanded. The thin film refers to U/W.\label{figure6}}
\end{figure}

In contrast, with increasing pressure, i.e. a compression of the $a_{U}$ axis, the anomaly in the phonons is suppressed \cite{Bouchet}, as experimentally verified in the bulk \cite{Raymond}. Thus, the CDW becomes weaker in the U/Nb films, as observed.

Furthermore, these anomalies result in a failure of the density-functional theory with quasi-harmonic thermodynamics to accurately predict the equation of state of $\alpha$-U at ambient conditions \cite{Dewaele}.

\section{Conclusions}

Our experiments have revealed that with uranium there are surprising differences in epitaxial relationships with different substrates and these drastically affect the subsequent behavior of the CDW at low temperature. It is for that reason we refer to the malleability of uranium. In the case of U/Nb films the $a_{U}$ axis is in slight compression and this leads to a reduction of the CDW amplitude as compared to the bulk, although little change in the transition temperature T0. In the case of U/W, where the epitaxial relationship is different from that found in U/Nb, the $a_{U}$ axis is in tension and, compared to the bulk, this increases (by almost a factor of three) the transition temperature of the CDW, as well as increasing its magnitude, and changing its form. These changes are consistent with the theory presented previously \cite{Bouchet,Raymond} for bulk $\alpha$-U, and emphasize the importance of the electron-phonon interaction.

Since the CDW is intimately connected to the superconductivity in bulk uranium \cite{Lander,Raymond}, we anticipate some interesting behavior in the U-films. In particular, for the U/W film the superconducting temperature should be suppressed for ambient pressure, but likely to be greater than the maximum of 2.3$\,$K found for bulk $\alpha$-U under pressure \cite{Lander,smith}, when the CDW is suppressed. Of course, such experiments under pressure with thin films are challenging \cite{Park}, and it is unknown at what pressure the strong CDW will be suppressed in such films.

The results reported here emphasize that new behavior may be expected when complex crystal structures (different from the well-known simple structures such as \textit{bcc}, \textit{fcc}, \textit{hcp}, and \textit{dhcp}) are used in epitaxial engineering. The only complex structure that has been examined previously is that of $\alpha$-Mn \cite{Grigorov}, but in this case numerous domains complicated the elucidation of new physics. No doubt the domain behavior can be complex, but by understanding the epitaxial relationships, such problems may be minimized. In the case of U/Nb we have one domain, and with U/W we were able to obtain two equally populated domains, both of which show identical behavior.

The light actinide elements (Pa, U, Np, and Pu) present new physics with their strong mixing of the 5\textit{f} and conduction states, and it seems likely that if simple crystal structures can be made by epitaxial engineering, then other consequences of the strong electron-phonon coupling, that should be intrinsically present in all these materials, may be found.

\bibliography{U_malleability}

\end{document}